# Low-energy electron holographic imaging of individual tobacco mosaic virions

*Jean-Nicolas Longchamp\*, Tatiana Latychevskaia, Conrad Escher, & Hans-Werner Fink*
*Physics Department, University of Zurich, Winterthurerstrasse 190, 8057 Zurich, Switzerland*

\*Corresponding Author
E-mail: longchamp@physik.uzh.ch

**Modern structural biology relies on NMR, X-ray crystallography and cryo-electron microscopy for gaining information on biomolecules at nanometer, sub-nanometer or atomic resolution. All these methods, however, require averaging over a vast ensemble of entities and hence knowledge on the conformational landscape of an individual particle is lost. Unfortunately, there are now strong indications that even X-ray free electron lasers will not be able to image individual molecules but will require nanocrystal samples.**
**Here, we show that non-destructive structural biology of single particles has now become possible by means of low-energy electron holography. As an example, individual tobacco mosaic virions deposited on ultraclean freestanding graphene are imaged at one nanometer resolution revealing structural details arising from the helical arrangement of the outer protein shell of the virus. Since low-energy electron holography is a lens-less technique and since electrons with a deBroglie wavelength of approximately 1 Angstrom do not impose radiation damage to biomolecules, the method has the potential for Angstrom resolution imaging of single biomolecules.**

Structural information about biomolecules at nanometer, sub-nanometer or atomic resolution is nowadays predominantly obtained by X-ray crystallography and NMR spectroscopy, whereby samples in the form of crystals or in liquids are studied. This, however, entails important structural information being averaged over many molecules. Thus, relevant details in molecules exhibiting diverse structural conformations remain undiscovered. Besides this drawback, these methods can only be applied to a limited subset of biological molecules that either readily crystallize for use in X-ray studies or are small enough for NMR investigations. A third approach towards imaging single particles is cryo-electron microscopy; however, for biological samples the possible resolution is limited by radiation damage caused by the high electron energy employed in conventional transmission electron microscopes (TEM)[1]. Due to the strong inelastic scattering of high-energy electrons, there is little hope for obtaining structural information at atomic resolution for a single entity. As the permissible dose is limited to $10 e^-/Å^2$ only, an individual molecule is destroyed long before an image of high enough quality could be acquired[2,3]. The radiation damage problem is usually circumvented by averaging over several thousand noisy images in order to attain a satisfactory signal-to-noise ratio[4]. The alignment and averaging routines inherent to high-resolution cryo-electron microscopy limit its application range to symmetric and particularly rigid objects, such as specific classes of viruses. Despite the shortcomings of the three conventional structural biology tools discussed above, one needs to express respect for the vast amount of data that has been generated over the past decades, reflected by the impressive volume of the current protein database.

Nevertheless, a milestone for structural biology would definitely be attained if methods and tools were available, that do away with averaging over an ensemble of molecules and enable structural biology on a truly single molecule level. To obtain atomic resolution information about the structure of any individual biological molecule, different concepts and technologies are needed. One approach of this kind is associated with the recent X-ray free electron laser (XFEL) projects. This approach initially appeared to be a promising method for gaining information from just one single biomolecule at the atomic scale by recording its X-ray diffraction pattern within the short time of just 10fs, before the molecule is decomposed by radiation damage. Unfortunately, there are now strong indications[5] that again averaging over a large number of molecules is inevitable in order to obtain images with a sufficiently high signal-to-noise ratio enabling numerical reconstruction of the diffraction pattern with atomic resolution[6–8].

The approach to structural biology at the single particle level described here, is motivated by the experimental evidence that low-energy electrons with a kinetic energy in the range of 50-250eV are harmless to biomolecules[9–11]. Even after exposing fragile molecules like DNA or proteins to a total electron dose of $10^6 e^-/Å^2$, i.e. more than five orders of magnitude higher than the critical dose in TEM, no radiation damage could be observed. This, combined with the fact that the deBroglie wavelengths associated with this energy range are between 0.7 and 1.7Å, makes low-energy electron microscopy an auspicious candidate for structural biology at the single molecule level[11,12]. During the last three decades, DNA, phages, viruses and individual ferritin proteins attached to carbon nanotubes were imaged by means of low-energy electron holography with nanometer resolution[9,10,13–15]. For imaging, the objects of interest used to be placed across bores in a membrane. Unfortunately, after such preparation the holographic record often suffered from biprism distortion limiting the resolution in the reconstruction[16]. This artifact is suppressed if the specimen is placed on an electrically conductive substrate with sufficient transparency for low-energy electrons[17–19]. Yet, the substrate has to be robust enough to withstand the deposition procedure[20]. It turned out that freestanding graphene, an atomically thin layer of carbon atoms arranged in a honeycomb lattice, fulfills all these requirements. Electron transmission measurements have shown that more than 70% of the low-energy electrons are transmitted through graphene and therefore are available for imaging objects deposited on the two dimensional substrate[18].

Tobacco mosaic virus (TMV) is a rod-shaped virus, approximately 300nm in length and 18nm in diameter, discovered at the end of the 19th century [21–24]. The first molecular model of TMV at atomic resolution was obtained from X-ray fiber diffraction experiments by Namba et al. in 1986[25,26]. Retrieving information about unstained TMV at the sub-nanometer scale is possible by means of cryo-electron microscopy since the 1980's[27,28]. The most recent models that can be found in the protein database are either obtained from X-ray fiber diffraction data (2.9Å resolution) or high-energy transmission electron microscopy investigations (5Å resolution). In both cases, the Angstrom resolution could only be obtained by averaging over a vast number of entities.

Here, we show that by means of low-energy electron holography, it is possible to image individual TMVs deposited on ultraclean freestanding graphene. The virions are imaged with one nanometer resolution exhibiting details of the helical TMV structure.

In our low-energy electron holographic setup (Fig. 1(a)), inspired by Gabor's original idea of in-line holography[29–31], a sharp (111)-oriented tungsten tip (Fig. 1(b)) acts as source of a divergent beam of highly coherent electrons[31–34]. The electron field emitter can be brought as close as 100nm to the sample with the help of a 3-axis nanopositioner. Part of the electron wave is elastically scattered off the object and hence is called the object wave, while the un-scattered part of the wave represents the reference wave. At a distant detector, the hologram, i.e. the pattern resulting from the interference of these two waves is recorded. The magnification of the imaging system is given by the ratio between detector-to-source distance and sample-to-source distance and can be as high as $10^6$. A hologram, in contrast to a diffraction pattern, contains the phase information of the object wave, and the object structure can thus be reconstructed unambiguously. The numerical reconstruction from the hologram is essentially achieved by back propagation to the object plane, which corresponds to evaluating the Fresnel-Kirchhoff integral transformation[35–40]. In low-energy electron holography, a lens-less technique not suffering from lens aberrations, the resolution limit is given by the deBroglie wavelengths ($\lambda$) and by the numerical aperture (NA) of the detector system. With $\lambda$ being as small as 0.7Å and NA=0.54, Angstrom and even atomic resolution shall eventually be possible.

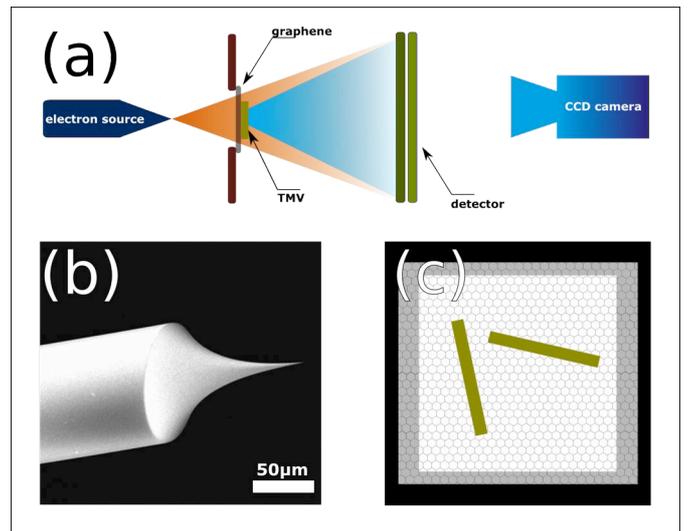

FIG. 1: Illustration of the experimental scheme for low-energy electron holography. (a) The source-to-sample distance amounts to typically 100-1000nm which leads to kinetic electron energies in the range of 50-250eV and the sample-to-detector distance is 70mm. With an electron detector of 75mm in diameter an acceptance angle of 56° is achieved (full angle). (b) SEM image of an electrochemically etched W(111) tip acting as field-emitter of a divergent coherent low-energy electron beam. (c) Schematic illustration of the sample geometry with two TMVs lying on ultraclean freestanding graphene suspended over a square aperture.

Ultraclean freestanding graphene, covering ion milled square-like apertures of approximately 500nm side length, is prepared by the platinum-metal catalysis method, described in detail recently elsewhere[41]. This method leads to large atomically clean areas, up to several square microns in size[19]. In Fig. 2(a), a low-energy electron transmission image of ultraclean freestanding graphene layer is presented. Transmission measurements have shown that single layer graphene exhibits a transparency of more than 70% for low-energy electrons[18]. Fig. 2(b) and 2(c) display low-energy electron holograms recorded at kinetic energies of 131eV and 125eV respectively. In these images, the rod-like virions deposited on graphene are apparent besides traces of contaminations resulting from the sample preparation procedure. Nevertheless, the cleanliness of the graphene is maintained throughout the entire preparation process to a sufficient level for imaging. A detailed description of the preparation and deposition method of TMV on graphene can be found in the supplementary material[42].

In Fig. 3(a) and 3(c), two high magnification holograms of TMV are displayed. From these holograms the shape of the corresponding virions is reconstructed at one nanometer resolution (see Fig. 3(b) and 3(d)). The diameter of the virion corresponds to 18nm as expected.

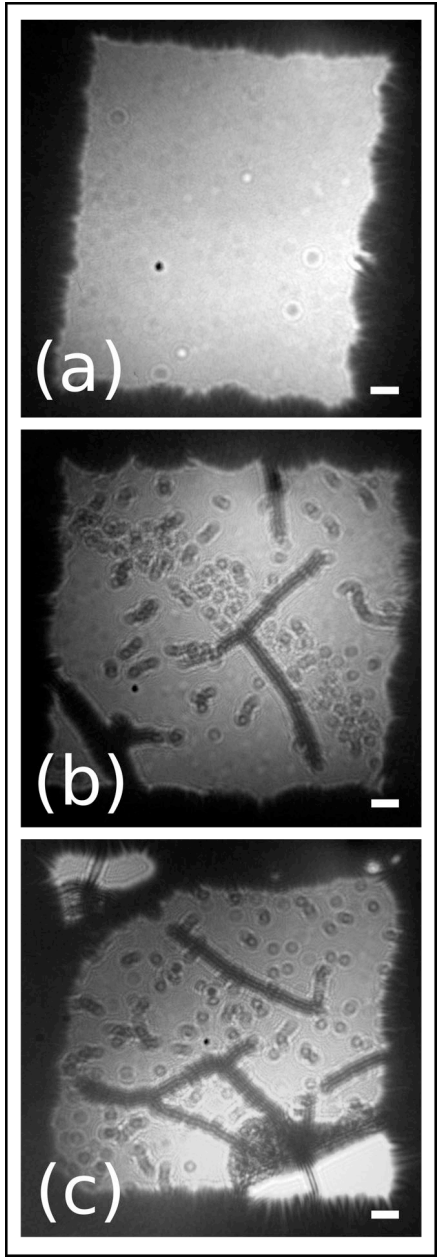

FIG. 2: Low-energy electron holograms before and after TMV deposition on ultraclean freestanding graphene. (a) Freestanding graphene covering a square aperture milled in a Pd-coated SiN membrane. Before TMV deposition the graphene layer is ultraclean. (b) and (c) Low-energy electron holograms of TMVs deposited onto freestanding graphene. The scale bar corresponds to 50nm.

Furthermore, one can observe, as emphasized by yellow arrows in Fig. 3(b), apex-like features on the rim of the virion, which we attribute to the helical arrangement of the outer protein shell of TMV. The spatial resolution attained in a hologram can be estimated using the Abbe criterion[43,44] and by measuring the largest angle under which interference fringes are observable[39,40,45]. This is illustrated in Fig. 4(b), where an intensity profile along the blue line in the hologram is displayed. The highest order interference fringe observed in the hologram is found at a scattering angle of 88mrad. Given the electron wavelength of 1.37Å for an electron kinetic

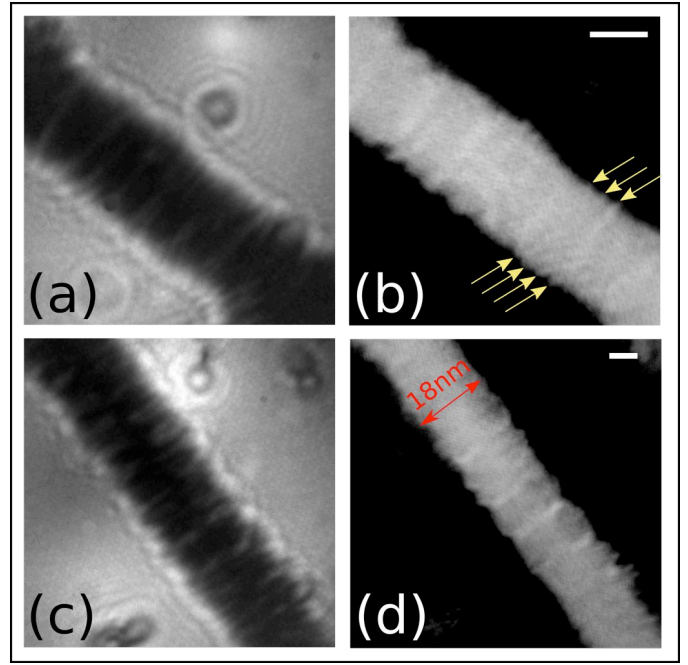

FIG. 3: High-magnification holograms of TMV and the respective reconstructions. (a) and (c) Holograms of TMV recorded with 80eV respectively 89eV electron energy. (b) and (d) reconstructed TMV images from (a) and (c) (inverted gray scale). Yellow arrows emphasize in (b) the presence of apex-like features on the rim of the TMV. These details are attributed to the helical structure of the virus. The scale bar corresponds to 10nm.

energy of 80eV, this angle corresponds to a spatial resolution of approximately 0.8nm.

Electrons with a kinetic energy of 80eV and 89eV exhibit deBroglie wavelengths of 1.37Å and 1.29Å respectively. This combined with the fact that in low-energy electron holography the resolution is neither limited by lens aberrations nor by radiation damage, Angstrom or even atomic resolution may be expected. The nanometer resolution that we report here has to be attributed to residual mechanical vibrations. In a low-energy electron hologram the spacing between consecutive interference fringes gradually decreases towards higher orders. Hence, high-order interference fringes and consequently high-resolution structural details are most susceptible to mechanical vibrations.

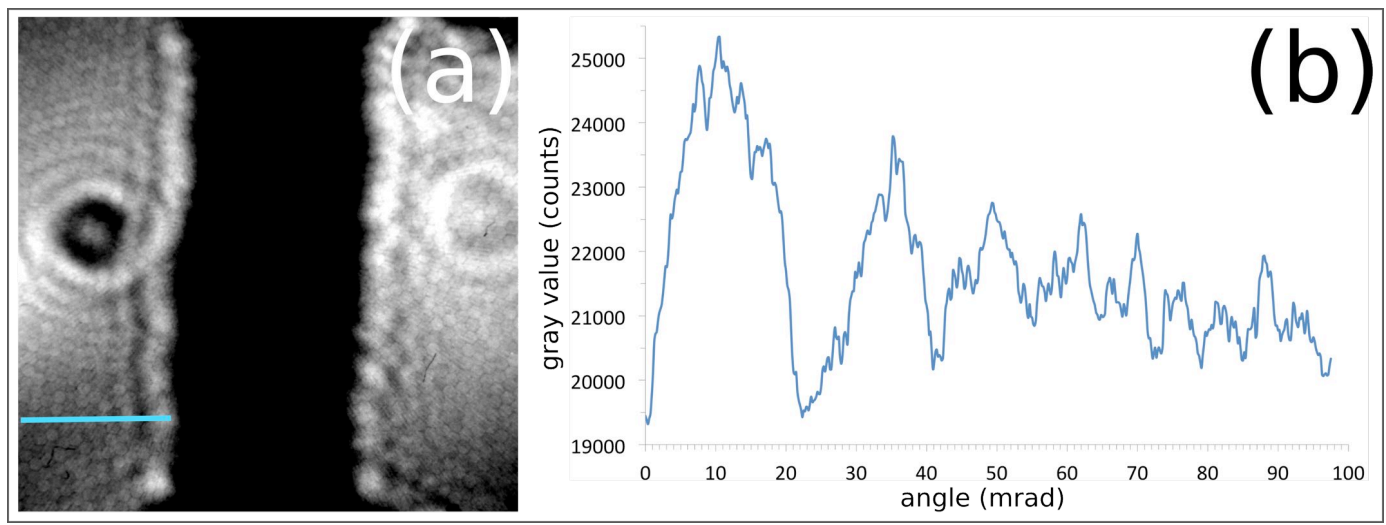

FIG. 4: Illustration of the resolution in a low-energy electron hologram of TMV. (a) 80 eV hologram of TMV presented in Fig. 3(a) but rotated and with enhanced contrast to reveal higher-order interference fringes. The hexagonal pattern observed in this image is due to the geometrical arrangement of the channels in the MCP-detector. (b) intensity scan along the blue line in (a). Interference fringes can be observed up to an angle of 88mrad. With a wavelength of 1.37Å, associated to 80eV electrons, this angle corresponds to a resolution of 0.8nm. The modulations due to the presence of the MCP channels are much lower than the signal obtained from the interference fringes of the hologram.

Even if the current resolution is of the order of one nanometer, one can already compare the images obtained with low-energy electrons with an atomic TMV model constructed by Jean-Yves Sgro[46] with the atomic coordinates available from the protein database (pdb id: 2tmv) (Fig. 5(a)). Fig. 5(b) and 5(c) are close-ups of the TMV images displayed in Fig. 3. Once the atomic TMV model is superimposed to match the width of the viruses (see Fig. 5(e) and 5(f)), the apex-like features that we previously attributed to the helical structure of the virus are now coinciding with the peaks of the helical structure in the model. The agreement between model and experimental data obtained from a single particle is remarkable. To take into account the kink in the TMV shown in Fig. 5(e), two copies of the model, rotated by 6° with respect to each other, are superimposed on the low-energy electron images. By this, further details, marked by yellow arrows in Fig. 5(b), which correspond to the helical structure of the virus are now congruent with the atomic model.

In Fig. 5(d), an intensity plot along the blue line present in Fig. 5(b) is displayed. The distance between depletions in this graph corresponds to approximately 2.35-2.40nm, a length that almost perfectly fits the expected 2.30nm helical pitch from X-ray fiber diffraction investigations[47,48]. The spatial resolution in the reconstructed TMV structure can be estimated by measuring the edge response[49]. The apex like features of the TMV structure represent an edge on which such a measurement can be done. By applying the common 10-90% limits (not shown here) on the intensity line scan in Fig. 5(d), a resolution of 0.95nm is obtained, in good agreement with the interference resolution criterion discussed above.

A further observation in Fig. 5(b-c) is the presence of bright stripes across the virus separated by 7nm. While these bright stripes correspond to higher material density, we do not understand the underlying biological origin. Nevertheless, we would like to hint that the observed 7nm distance is very close to the literature value of 6.9nm associated with the thickness of a TMV subunit[48,51]. Similar features have been observed in TEM investigations of uranyl acetate stained in-vitro assembled viruses[52,53] but, to our knowledge, never on in-vivo purified TMV.

In this letter, we report the first nanometer resolved images of single tobacco mosaic virions obtained from low-energy electron investigations. Details revealed on the rim of the TMV are attributed to the helical structure of the virus by confronting our images with an atomic model based on the coordinates available from the protein database. With this, we have demonstrated the potential of low-energy electron holography for structural biology at the single particle level.

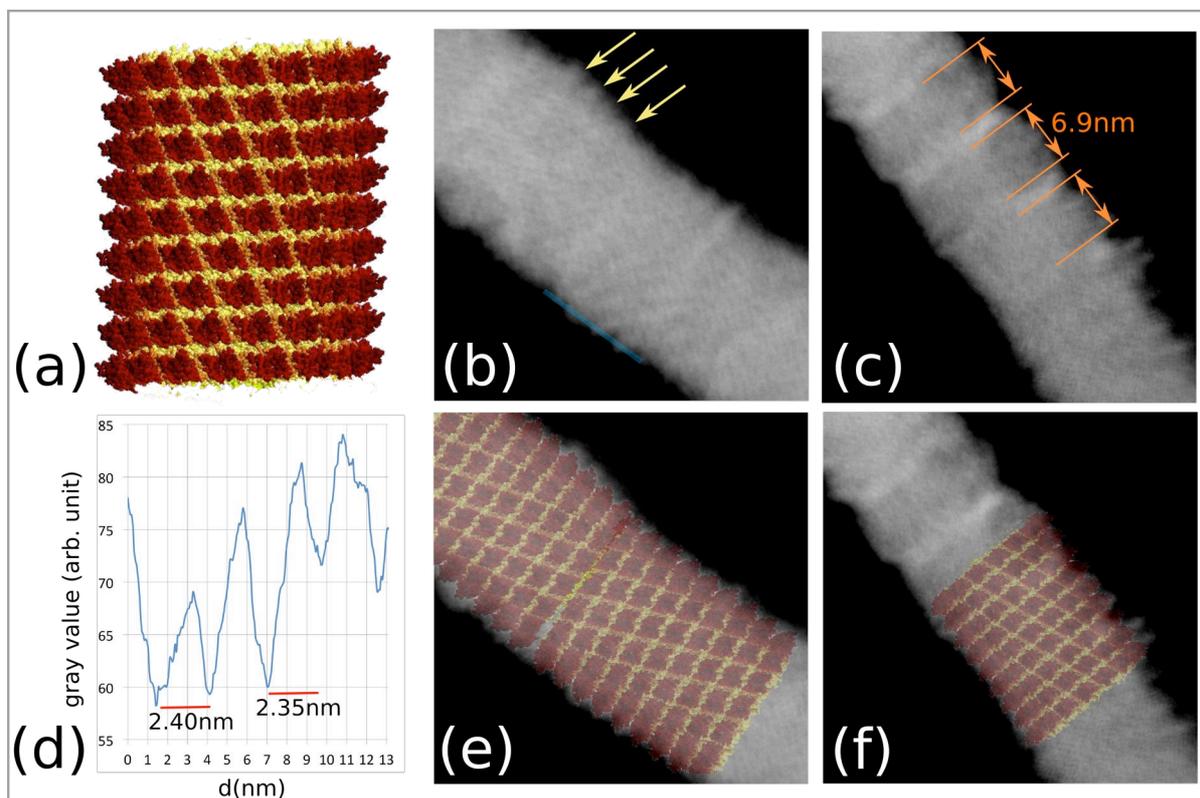

FIG. 5: Comparison of an atomic TMV model with the low-energy electron images. (a) Atomic TMV model[46,50] constructed from the coordinates available from the protein database (pdb id: 2tmv). (b-c) Close-ups of the reconstructed TMV images presented in Fig. 3. In (b) yellow arrows mark further apex-like features coinciding with the atomic model. In (c) the distance between bright stripes is emphasized and marked according to the literature value of 6.9nm for the thickness of a subunit. (d) Intensity scan along the blue line in (b). (e-f) Atomic TMV model superimposed on the same images as in (b-c).

The current nanometer resolution will be pushed to angstrom resolution in the near future by improving the mechanical stability of the microscope. Furthermore, we have recently reported that by employing a slightly modified experimental setup, where a parallel beam of low-energy electrons is illuminating the sample, we could image a region of 210nm in diameter of freestanding graphene with 2Å resolution[54]. This experimental scheme will now also be used to image TMV at similar resolution.

The authors are grateful for financial support by the Swiss National Science Foundation (grant number PZ00P2 148084). We would like to thank Annette Niehl and Manfred Heinlein from the CNRS Strasbourg for the TMV purification and helpful discussions. We also would like to thank Kishan Thodkar and Christian Schönenberger from the University of Basel for providing us with their CVD grown graphene.

**Supplementary Information - Sample Preparation**

Ultraclean freestanding graphene, covering ion milled square-like apertures of approximately 500nm side length, is prepared by the platinum-metal catalysis method[1]. Thereafter, the cleanliness of the as-prepared graphene is inspected in our low-energy electron holography microscope operating under UHV conditions. For TMV deposition, graphene samples are then taken out of the low-energy electron microscope and under ambient conditions, a drop of a 0.5ng/µl aqueous TMV solution is subsequently applied onto the graphene. A few seconds were given for the viruses to sediment before the excess water was removed by blotting paper. In order to achieve a sufficiently clean sample for low-energy electron investigations, the sample is kept for 30min at a temperature of at least 50°C in air prior to the re-introduction into the electron microscope.

Virions were prepared by following the recipe published by Niehl et al.[2]. They were extracted from TMV-infected nicotiana benthamiana leaves, purified by precipitation and sedimentation and re-suspended in 10mM sodium-phosphate buffer, pH 7.4 and at a final concentration of 0.5ng/ml.

We are aware, that the deposition method used here is only applicable for a very small subset of biological entities that withstand the moderate heat treatment. However, we have recently shown that it is possible to electrospray deposit *in vacuo* gold nanorods of 20MDa molecular weight onto freestanding graphene without damaging the atomically thin substrate[3]. It demonstrates that *in vacuo* deposition of proteins onto graphene is within arm's reach.